\documentstyle[epsf,12pt]{article}

\begin{document}
\mbox{ }
\rightline{UCT-TP-260/01}\\
\rightline{MZ-TH/01-11}\\
\rightline{April 2001}\\
\vspace{3.5cm}
\begin{center}
{\Large \bf Ratio of strange to non-strange quark condensates in  QCD}
\footnote{Work supported
in part by the Volkswagen Foundation}\\
\vspace{.5cm}
{\bf C. A. Dominguez $^{(a)}$, A. Ramlakan $^{(a)}$,
K. Schilcher $^{(b)}$}\\[.5cm]
$^{(a)}$Institute of Theoretical Physics and Astrophysics\\
University of Cape Town, Rondebosch 7700, South Africa\\

$^{(b)}$ Institut f\"{u}r Physik, Johannes Gutenberg-Universit\"{a}t,\\
Staudingerweg 7, D-55099 Mainz, Germany\\ [.5cm]
\end{center}
\vspace{.5cm}
\begin{abstract}
\noindent
Laplace transform QCD sum rules for two-point functions related to
the strangeness-changing scalar and pseudoscalar Green's functions
$\psi(Q^2)$ and $\psi_5(Q^2)$,
are used to determine the subtraction constants $\psi(0)$ and $\psi_5(0)$,
which fix the ratio $R_{su}\equiv \frac{<\bar{s}s>}{<\bar{u}u>}$. Our
results are $\psi(0)= - (1.06 \pm 0.21) \times 10^{-3} \; \mbox{GeV}^4$,
$\psi_5(0)=  (3.35 \pm 0.25) \times 10^{-3} \; \mbox{GeV}^4$,  and
$R_{su}\equiv \frac{<\bar{s}s>}{<\bar{u}u>} = 0.5 \pm 0.1$.
This implies corrections to kaon-PCAC at the level of $ 50 \%$, which
although large, are not inconsistent with  the size of the corrections to
Goldberger-Treiman relations in $SU(3)\otimes SU(3)$.

\end{abstract}
\newpage
\setlength{\baselineskip}{1.5\baselineskip}
\noindent
The quark vacuum condensate ratio $R_{su}\equiv
\frac{<\bar{s}s>}{<\bar{u}u>}$, with
$<\bar{u}u>\, \simeq <\bar{d}d>$, is an important quantity which measures
the size of flavour SU(3) symmetry breaking in the QCD vacuum \cite{P}.
It also enters in a variety of QCD sum rules \cite{SVZ}, particularly
in those used to determine baryon masses \cite{DJN}. In addition, it
provides a measure of the corrections to kaon PCAC (Partial Conservation
of the Axial-Vector Current), provided the quark masses are known and
neglecting corrections to pion PCAC \cite{P},\cite{B}.
In fact, let us consider the two-point function involving the
flavour-changing axial-vector current divergences
\begin{equation}
\psi _5( q^2) =i \int d^4x\ e^{iqx}\left\langle 0| T (
\partial ^\mu A_\mu ( x)\; \partial ^\nu A_\nu ^{\dagger }(
0) ) |0 \right\rangle \ ,  
\end{equation}
where flavour indices have been omitted for simplicity, and 
\begin{equation}
\partial^\mu A_\mu(x)|^{j}_{i} = (m_i + m_j) :\bar{\psi_j}(x)\;
 i \gamma_5 \;
\psi_i(x):  \; ,
\end{equation}
and (i,j) are flavour indices. A well known Ward identity fixes the value
of $\psi_5(q^2)$ at zero momentum \cite{P},\cite{B}, viz.
\begin{equation}
\psi _5(0)|_i^j =-( m_i+m_j)
\left\langle \overline{\psi }_j\psi _j+\overline{\psi }_i\psi
_i\right\rangle \; ,
\end{equation}
which is a renormalization group invariant quantity. Saturation of
Eq.(1) with the lowest lying pseudoscalar meson then leads to the
Gell-Mann, Oakes, Renner relations \cite{P}
\begin{equation}
-( m_u+m_d)
\left\langle \overline{u} u+\overline{d}d \right\rangle
= 2 f_\pi^2 \mu_\pi^2\; ,
\end{equation}
for (i,j)=(u,d), and
\begin{equation}
-( m_s+m_u)
\left\langle \overline{s} s+\overline{u}u \right\rangle
= 2 f_K^2 \mu_K^2\; ,
\end{equation}
where now (i,j)=(u,s),
$f_\pi = 92.4 \pm 0.4 \;\mbox{MeV}$, and $f_K = 113.0 \pm 1.5\;\mbox{MeV}$
\cite{PDG}. Considering Eq.(3) for up- and strange-quark flavours and
then for up- and down-quark flavours and taking the ratio leads to
\begin{equation}
R_{AA} \equiv \frac {\psi_5(0)|^s_u} {\psi_5(0)|^d_u}
= \frac{1}{2} \;
\frac{m_s+m_u}{m_u+m_d} \;(1 + \frac{<\bar{s}s>}{<\bar{u}u>}) \; ,
\end{equation}
where $<\bar{u}u>\, \simeq <\bar{d}d>$ is a very good approximation in this
case \cite{DR}. In principle, this relation could be used to determine
$R_{su}$, if the ratio of the subtraction constants
$\psi_{5}(0)$, i.e. $R_{AA}$, is known independently. Using current
values of the quark masses \cite{PDG} in Eq.(6) leads to
\begin{equation}
R_{su}\equiv \frac{<\bar{s}s>}{<\bar{u}u>} \simeq 0.15\; R_{AA} - 1 \; .
\end{equation}
Since $R_{AA}$ is expected to be of order $\cal{O}$$(10)$ (from
using PCAC), this procedure would lead to very large uncertainties,
unless $\psi_{5}(0)$ were to be determined with extreme accuracy.
An alternative
method not affected by this problem was first proposed in \cite{DL}, and
it is based on examining $\psi_{5}(0)$ together with $\psi(0)$, where
\begin{equation}
\psi ( q^2) =i \int d^4x\ e^{iqx}\left\langle 0| T (
\partial ^\mu V_\mu ( x) \;\partial ^\nu V_\nu ^{\dagger }(
0) ) |0 \right\rangle \ ,  
\end{equation}
with
\begin{equation}
\partial^\mu V_\mu(x)|^{j}_{i} = (m_j - m_i) :\bar{\psi_j}(x)\;i  \
\psi_i(x):  \; ,
\end{equation}
being the flavour-changing vector current divergence. The Ward identity
analogous to Eq.(3) is now
\begin{equation}
\psi (0) _i^j =-( m_j-m_i)
\left\langle \overline{\psi }_j\psi _j-\overline{\psi }_i\psi
_i\right\rangle \; .
\end{equation}
Defining the ratio
\begin{equation}
R_{VA}\equiv \frac{\psi( 0) _u^s}{\psi _5(0) _u^s}\ ,
\end{equation}
it follows that
\begin{equation}
R_{su} \equiv
\frac{\left\langle \overline{s}s\right\rangle }{\left\langle \overline{u}
u\right\rangle }\simeq \frac{1+R_{VA}}{1-R_{VA}} \; ,
\end{equation}
where the reasonable approximation $(m_s-m_u) \simeq (m_s+m_u)$ has been
made. This procedure can provide a more accurate determination of
$R_{su}$, as discussed in \cite{DL}. Since then,
considerable progress has been made on the QCD calculation of the
two-point functions
Eqs.(1) and (8), viz. the perturbative expressions are now known up to
four loops \cite{PQCD}, quark-mass corrections have been calculated
up to order $\cal{O}$$(m_q^6)$, and the issue of quark-mass logarithmic
singularities has been satisfactorily resolved \cite{CHs}-\cite{JM}. On
the other hand, there is now better experimental information on the
hadronic spectral functions in both the scalar \cite{S} and the
pseudoscalar channels \cite{PDG}. In view of this, we discuss here a new
determination of the subtraction constants $\psi_{(5)}(0)$ which are then
used to determine the ratio $R_{su}$ through
Eq.(12).\\

In order to relate the constants $\psi_{(5)}(0)$ to the two-point
functions $\psi_{(5)}(q^2)$, Eqs.(1) and (8), one defines the
auxiliary functions
\begin{equation}
q^2 D_{(5)}(q^2) =\psi_{(5)}( q^2) -\psi_{(5)} (0)\; .
\end{equation}
The derivatives of these functions, i.e.
\begin{equation}
\xi(q^2) = \frac{\partial}{\partial q^2} D(q^2) \; ,
\end{equation}
\begin{equation}
\phi(q^2) = \frac{\partial}{\partial q^2} D_5(q^2) \; ,
\end{equation}
have the same imaginary part as $\psi_{(5)}(q^2)$, and
satisfy the following dispersion relations
\begin{equation}
\left.
\begin{array}{c}
\xi ( Q^2) \\
\phi ( Q^2)
\end{array}
\right\}
=\frac {1}{\pi} \int_{0}^{\infty} \frac{ds}{s} \;
\frac{Im \;\psi_{(5)}( s)}{( s+Q^2)^2}\ \;, 
\end{equation}
where $Q^2 \equiv - q^2 \geq 0$. 
After Laplace improvement, these dispersion relations become
\begin{equation}
\left. 
\begin{array}{c}
\xi \left( M_L^2\right) \\ 
\phi \left( M_L^2\right)
\end{array}
\right\} =   \frac {1}{\pi} \int_{0}^{\infty} \frac{ds}{s}\;
e^{-s/M_L^2} \;
Im \;\psi _{(5)}(s) \ \;.
\end{equation}
According to quark-hadron duality, the
left hand side of Eq.(17) is computed in QCD using the Operator Product
Expansion to organize the perturbative series and the power corrections,
and the right hand side is saturated with the experimental data.
Using the available QCD expressions for $\psi_{(5)}(q^2)$
\cite{CHs}-\cite{JM}, performing the derivatives, Laplace transforming,
and after Renormalization Group improvement,
we obtain the following QCD representations for the left hand sides of
Eq.(17) for three quark flavours
\newpage
\begin{eqnarray}
\left. 
\begin{array}{c}
\xi \left( M_L^2\right) \\ 
\phi \left( M_L^2\right)
\end{array}
\right\} &=&\frac 3{8\pi ^2}\frac{\overline{m}_s^2\left( M_L^2\right) }{M_L^2
}\left\{ 1+\frac{\overline{\alpha }_s\left( M_L^2\right) }{\pi} \right.
\left( \frac{17}{3}-2 \;\Psi \left( 1\right) \right)  \nonumber \\
&&\ \ +\left( \frac{\overline{\alpha }_s\left( M_L^2\right) }{\pi} \right)
^2\left[  \frac{9631}{144}-\frac{35}{2} \;
\zeta \left( 3\right) -\frac{95
}{3}\;\Psi \left( 1\right) \right. \nonumber \\
&&\ \ \left. +\frac{17}{4}\left( \Psi ^2\left( 1\right) -\Psi ^{^{\prime }}
\left(
1\right) \right) \allowbreak \allowbreak \allowbreak  \right] 
\nonumber \\
&&\ \ +\left( \frac{\overline{\alpha }_s\left( M_L^2\right) }{\pi} \right)
^3\left[ \frac{4748953}{5184}+\frac{715}{12}\;\zeta \left( 5\right) -
\frac{91519}{216}\;\zeta \left( 3\right) -\frac{\pi ^4}{36}
\right. \nonumber \\
&&\ \ -\allowbreak 2\left( \frac{4781}{18}-\frac{475}{8}
\;\zeta \left( 3\right)
\right) \Psi \left( 1\right) +\frac{229}{2}\left( \Psi ^2\left( 1\right) -
\Psi
^{^{\prime }}\left( 1\right) \right)  \nonumber \\
&& \ \  
 +\left. \frac{221}{24} \left.
 \left( 3\;\Psi \left( 1\right) \Psi ^{^{\prime
}}\left( 1\right) -\Psi ^3\left( 1\right) -\Psi ^{^{\prime \prime }}\left(
1\right) \right)  \allowbreak 
  \right] \right\}  \nonumber \\
&&\ \ +\frac {3}{4\pi ^2}\frac{\overline{m}_s^4\left( M_L^2\right) }{M_L^4}
 \left\{  2-\Psi \left( 2\right) 
 +\frac{\overline{\alpha }_s\left( M_L^2\right) }{\pi} \right.
 \left[ \frac{41}
{3}-4\;\zeta \left( 3\right)   \right. \nonumber \\
&&\ \ 
 - \left. \frac{28}{3}\; \Psi \left( 2\right) \left. +2\left( \Psi
^2\left( 2\right) -\Psi ^{^{\prime }}\left( 2\right) \right) \right]
\right\}  \nonumber \\
&&\ \ -\frac {1}{2}\frac{\overline{m}_s^2\left( M_L^2\right) }{M_L^6}
\left\{
 \left\langle m_s\overline{s}s\right\rangle \left[ 1+\frac{\overline{\alpha}
_s\left( M_L^2\right) }{\pi} \left( \frac{14}{3}-2
\;\Psi \left( 3\right) \right)
\right] \right.  \nonumber \\
&&\ \ \pm 2\left\langle m_s\overline{u}u\right\rangle \left[ 1+
\left.\frac{
\overline{\alpha }_s\left( M_L^2\right) }{\pi}
 \left( \frac{17}{3}-2\;\Psi \left(
3\right) \right) \right] \right\}  \nonumber \\
&&\ \ -\frac 18\frac{\overline{m}_s^2\left( M_L^2\right) }{M_L^6}
\left\langle \frac{\alpha _s}\pi G^2\right\rangle -\frac 3{8\pi ^2}\frac{
\overline{m}_s^6\left( M_L^2\right) }{M_L^6}\Psi \left( 3\right) +\frac{\psi
_{\left( 5\right) }\left( 0\right) }{M_L^4}\ ,  
\end{eqnarray}
where $\Psi(z)$ is the di-gamma function, primes stand for its
derivatives, and $\zeta(z)$ is Riemann's zeta function. The four-quark
vacuum condensate term of dimension-six has been omitted above as
it makes a negligible contribution, and the up-quark mass has been
neglected in comparison with $m_s$. The four-loop expressions for the
running coupling and quark mass in the $\overline{MS}$ renormalization
scheme, and for three flavours, are \cite{PQCD}
\begin{eqnarray}
\frac{\overline{\alpha }_s\left( Q^2\right) }\pi &=&\frac 49\frac 1L-\frac{
256}{729}\frac{LL}{L^2}  \nonumber \\
&&+\left[ 6794-16384\left( LL-LL^2\right) \right] \frac 1{59049}\frac 1{L^3}+
{\cal O}\left( \frac {1}{L^4}\right) \ ,
\end{eqnarray}
\newpage
\begin{eqnarray}
\overline{m}_j\left( Q^2\right) &=&\frac{\widehat{m}_j}{\left( \frac
12L\right) ^{\frac 49}}\left\{ 1+\left( 290-256LL\right) \frac 1{729}\frac
1L\right.  \nonumber \\
&&\nonumber \\
&&+ \left[  \frac{550435}{1062882}-\frac{80}{729}\;\zeta \left(
3\right) \right. \nonumber \\
&&\nonumber \\
&&-\left( 388736LL-106496LL^2\right)\left.  \frac 1{531441} \right] \frac
1{L^2}  \nonumber \\
&&\nonumber \\
&&+\left[  -\frac{126940037}{1162261467}-\frac{256}{177147}\;\beta _4+
\frac{128}{19683}\;\gamma _4+\frac{7520}{531441}\;\zeta \left( 3\right) 
\right. \nonumber \\
&&\nonumber \\
&& + \left( -\frac{611418176}{387420489}+
\frac{112640}{531441}\;\zeta \left(
3\right) \right) LL+\frac{335011840}{387420489}LL^2  \nonumber \\
&&\nonumber \\
&&- \left. \frac{149946368}{1162261467}LL^3  \right] \frac 1{L^3}
\left. +
{\cal O}\left( \frac {1}{L^4}\right) \right\} \ ,  
\end{eqnarray}

where $L = \ln (Q^2/\Lambda_{QCD}^2)$, $LL = \ln L$, and
\begin{equation}
\beta _4=-\frac{281198}{4608}-\frac{890}{32}\;\zeta \left( 3\right) \ , 
\end{equation}
with $\gamma_4 = 88.5258$ \cite{G4}, and $\widehat{m}_j$ is the invariant
quark-mass.
The terms of order ${\cal O}$$\left( \frac {1}{L^4}\right) $ above
are known up to a constant not determined by the renormalization group.
This constant can be estimated e.g. using Pad\`{e}
approximants \cite{K3}. However, we have checked that our final results
are essentially insensitive to terms of this order.\\

Before we proceed to discuss the hadronic parametrization of the scalar
and pseudoscalar spectral functions, it should be stressed that these
spectral functions are also used to determine the strange-quark mass
from QCD sum rules for $\psi_{5}(q^2)$ \cite{MS}
and $\psi(q^2)$ \cite{CHs}-\cite{JM},\cite{IT}.
The fact that $m_s$ is one of the most important
parameters in Eq.(18), and given the wide range of values obtained
from QCD sum rules \cite{REV}, it is imperative to achieve consistency
between the results for $m_s$ from the scalar and the pseudoscalar
channels. Fully consistent results are obtained as explained below. Another
very important parameter in Eq.(18) is $\Lambda_{QCD}$. The value
extracted from a variety of experimental data has been increasing steadily
over the years, from $\Lambda_{QCD} \simeq 100 - 200 \;\mbox{MeV}$
at the birth of the QCD sum rules \cite{SVZ}, to something as high as
$\Lambda_{QCD} (N_f = 3) \simeq 300 - 450 \;\mbox{MeV}$ lately \cite{PDG}.
Such high values make  radiative corrections to Green's functions
the  overwhelming terms in the Operator Product Expansion, thus
reducing the numerical importance of the power corrections as parametrized
by quark and gluon vacuum condensates. In fact, it has been argued in
\cite{C2} that if $\Lambda_{QCD} (N_f = 3)$ exceeds
$ \simeq 330 \;\mbox{MeV}$
the QCD sum rule program may break down, and it becomes extremely 
difficult to extract numerical values for the condensates from the data (see also
\cite{BI}).
As for the other QCD parameters in Eq.(18), the gluon condensate has the
value \cite{GC}
$\left\langle \frac{\alpha _s}{\pi} G^2\right\rangle \simeq 0.01-0.04 \
\mbox{GeV}^4$. The light-quark condensate (at 1 GeV) is
$\left\langle \overline{u}u\right\rangle \simeq - 10^{-2} \;\mbox{GeV}^3$.
For the strange-quark condensate, since it is the object of this
determination, an iteration procedure of fast convergence has been adopted,
starting from $\left\langle \overline{s}s\right\rangle \simeq
\left\langle \overline{u}u\right\rangle $.\\

Turning to the hadronic sector, the spectral function in the pseudoscalar
channel is parametrized as in \cite{MS}, viz.
\begin{eqnarray}
\frac {1}{\pi} Im\; \psi _5\left( s\right) |_{HAD}
&=&2f_K^2M_K^4\delta \left( s-M_K^2\right)  \nonumber \\
&&+\frac {1}{\pi} Im \psi _5\left( s\right) |_{K\pi \pi
}\left[ \frac{BW_1\left( s\right) +\lambda
BW_2\left( s\right) }{(1+\lambda )
}\right] \ ,  
\end{eqnarray}
where the threshold behaviour of the resonant piece,
determined using chiral perturbation theory, is given by 
\begin{equation}
\frac {1}{\pi} Im \;\psi _5\left( s\right) |_{K\pi \pi }=
\frac{M_K^4}{2f_\pi ^2} \;\frac {3}{2^8\pi ^4}\;
\frac{I\left( s\right) }{s\left(
M_K^2-s\right) ^2} \;\theta \left( s-M_K^2\right) \ ,  
\end{equation}
with
\begin{eqnarray}
I\left( s\right) &=&\int_{M_K^2}^{s}\frac{du}{u}\left( u-M_K^2\right)
\left( s-u\right) \times \left\{ \left( M_K^2-s\right) \left( u-
\frac{\left( s+M_K^2\right) }{2}\right)  \right. \nonumber \\
&& \left.
-\frac {1}{8u}\left( u^2-M_K^4\right) \left( s-u\right) +\frac{3}{4}\left(
u-M_K^2\right) ^2 | F_{K^{*}}\left( u\right) | ^2   \right\} \; ,
\end{eqnarray}
where
\begin{equation}
\left| F_{K^{*}}\left( u\right) \right| ^2=\frac{\left(
M_{K^{*}}^2-M_K^2\right) ^2+M_{K^{*}}^2\Gamma _{K^{*}}^2}{\left(
M_{K^{*}}^2-u\right) ^2+M_{K^{*}}^2\Gamma _{K^{*}}^2}\ ,  
\end{equation}
is the contribution from the resonant sub-channel $K^{*}(892)-\pi$.
The parameter$\lambda$ above controls the importance of the second radial
excitation (K(1830)) relative to the first (K(1460)) ($\lambda \simeq 1$),
with both resonances
being parametrized by  Breit-Wigner forms $BW_{1,2}(s)$.
Due to the approximation $m_u = 0$, the pion mass has been
neglected. There is another resonant sub-channel, the $\rho(770) K$,
which turns out to be numerically negligible \cite{MS}. As usual, this
resonant hadronic parametrization is used up to some threshold energy,
$s_{0A}$, the continuum threshold, after which the spectral function is
assumed to be given by perturbative QCD.\\
In the case
of the scalar channel, there is experimental data on  $K \pi$ phase
shifts \cite{S} which can be used to reconstruct the spectral function
given by
\begin{equation}
\frac {1}{\pi} Im\; \psi \left( s\right) =\frac {3}{32\pi ^2}
\frac{\sqrt{\left( s- s_{+}\right) \left( s- s_{-}
\right)}}{s} \; \left| d(s)\right|^2 \; ,
\end{equation}
where $s_{\pm} = (\mu_K \pm \mu_\pi)^2$, and $d(s)$ is the scalar form
factor. We have used the method of \cite{IT}, based on the Omn\`{e}s
representation, to relate $d(s)$ to the experimental phase shifts. In this
way, the results for $m_s$ from the pseudoscalar channel are fully
compatible with those from the scalar channel, giving an invariant
strange-quark mass
\begin{equation}
 \widehat{m}_s = 140 \pm 10 \;\mbox{MeV} \; ,
\end{equation}
for $\Lambda_{QCD}$ in the range $\Lambda_{QCD}\simeq 300-350\;
\mbox{MeV}$.
As mentioned earlier, higher values of $\Lambda_{QCD}$ may invalidate the QCD
sum rule program in general, and they do so in this particular application,
as they lead to serious instabilities in the results for $\psi_{(5)}(0)$.
Therefore, we restrict $\Lambda_{QCD}$ to the above range. In Figs. 1 and
2 we show typical results for $\psi_5(0)$ and $\psi(0)$, respectively,
obtained using the values $\Lambda_{QCD} = 300 \;\mbox{MeV}$,
$\widehat{m}_s = 145 \;\mbox{MeV}$,
$\left\langle \frac{\alpha _s}{\pi} G^2\right\rangle = 0.01 \; \mbox{GeV}^4$,
and the asymptotic freedom thresholds $s_{0A} = 6\; \mbox{GeV}^2$,
and $s_{0V} = 4 \;\mbox{GeV}^2$.
These results lead to the ratio $R_{su}$
shown in Fig.3. As it may be appreciated from these figures, the results
are nicely insensitive to changes in the Laplace variable $M_L^2$; they
are also reasonably stable against changes in the continuum
thersholds, $s_{0V,A}$.
Increasing $\Lambda_{QCD}$ leads to smaller values of
$\widehat{m}_s$, and higher values of the ratio $R_{su}$. After allowing
for variations in all the relevant parameters, within the ranges indicated
above, we obtain
\begin{equation}
\psi_5(0)=  (3.35 \pm 0.25) \times 10^{-3} \; \mbox{GeV}^4 \;,
\end{equation}
\begin{equation}
\psi(0)= - (1.06 \pm 0.21) \times 10^{-3} \; \mbox{GeV}^4 \;,
\end{equation}
\begin{equation}
R_{su}\equiv \frac{<\bar{s}s>}{<\bar{u}u>} =  0.5 \pm 0.1      \;.
\end{equation}
A comparison of these results with previous determinations based on
similar methods \cite{DL}, \cite{OLD} is not very
enlightening, as most of the
old analyses were done at the two-loop level in perturbative QCD, they
were affected by logarithmic quark-mass singularities, and did not make
use of the latest experimental data to reconstruct the hadronic spectral
functions. However, these problems do not affect determinations based
on different methods. For instance, from QCD sum rules for baryon masses
\cite{DJN} $R_{su} = 0.6 \pm 0.1$ has been obtained, while an estimate
based on non-perturbative quark-mass generation \cite{PRS} gives
$R_{su}\simeq 0.7$ (with no error estimates). Our result, Eq.(30) is
consistent with these values. Furthermore, Eq.(29) leads to a correction
to kaon-PCAC, $\delta_K$, defined as
\begin{equation}
\psi_5(0) =  2 f_K^2 \mu_K^2 (1-\delta_K) \; ,
\end{equation}
where using Eq.(29) one finds
\begin{equation}
\delta_K  \simeq 0.5 \;.
\end{equation}
This result points to a rather large correction to kaon-PCAC, although it is 
not inconsistent with
the expected size of the corrections to Goldberger-Treiman relations in
$SU(3)\otimes SU(3)$ \cite{GTR}.

\begin{center}
{\bf Figure Captions}
\end{center}

Figure 1. The subtraction constant $\psi_5(0)$ as a function of the
Laplace variable $M_L^2$, for $\Lambda_{QCD} = 300 \; \mbox{MeV}$,
$\widehat{m}_s = 145 \;\mbox{MeV}$,
$\left\langle \frac{\alpha _s}{\pi} G^2\right\rangle = 0.01 \; 
\mbox{GeV}^4$,
and the continuum thresholds $s_{0A} = 6\; \mbox{GeV}^2$,
and $s_{0V} = 4 \; \mbox{GeV}^2$.\\

Figure 2. The subtraction constant $\psi(0)$ as a function of the
Laplace variable $M_L^2$, for $\Lambda_{QCD} = 300 \; \mbox{MeV}$,
$\widehat{m}_s = 145 \;\mbox{MeV}$,
$\left\langle \frac{\alpha _s}{\pi} G^2\right\rangle = 0.01 \; 
\mbox{GeV}^4$,
and the continuum thresholds $s_{0A} = 6\; \mbox{GeV}^2$,
and $s_{0V} = 4 \;\mbox{GeV}^2$.\\

Figure 3. The ratio
$R_{su}\equiv \frac{<\bar{s}s>}{<\bar{u}u>}$ as a function of the
Laplace variable $M_L^2$, for $\Lambda_{QCD} = 300 \; \mbox{MeV}$,
$\widehat{m}_s = 145 \;\mbox{MeV}$,
$\left\langle \frac{\alpha _s}{\pi} G^2\right\rangle = 0.01 \; 
\mbox{GeV}^4$,
and the continuum thresholds $s_{0A} = 6\; \mbox{GeV}^2$,
and $s_{0V} = 4 \;\mbox{GeV}^2$.

\newpage
\begin{figure}[tp]
\epsffile{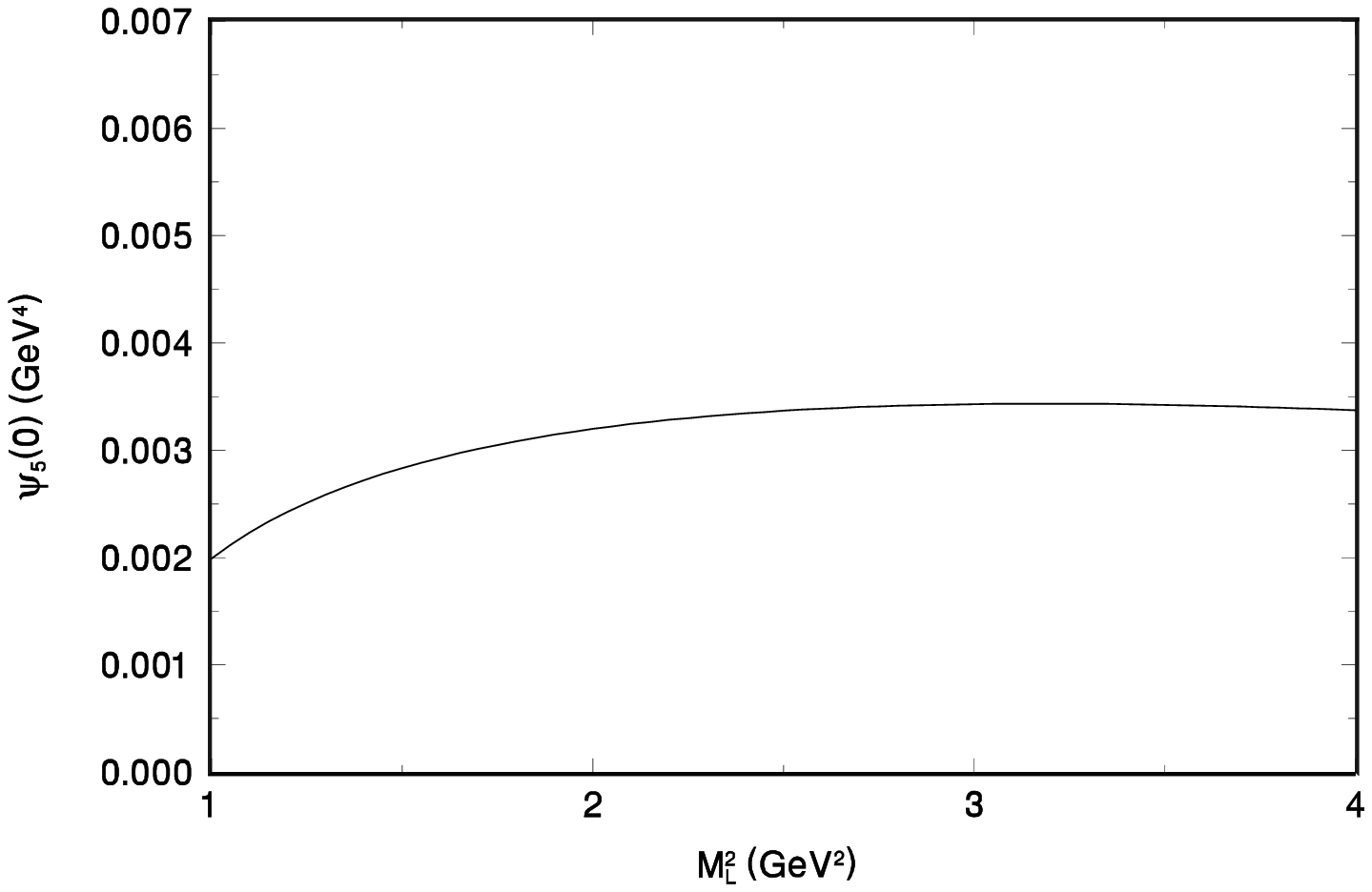}
\caption{}
\end{figure}
\newpage
\begin{figure}[tp]
\epsffile{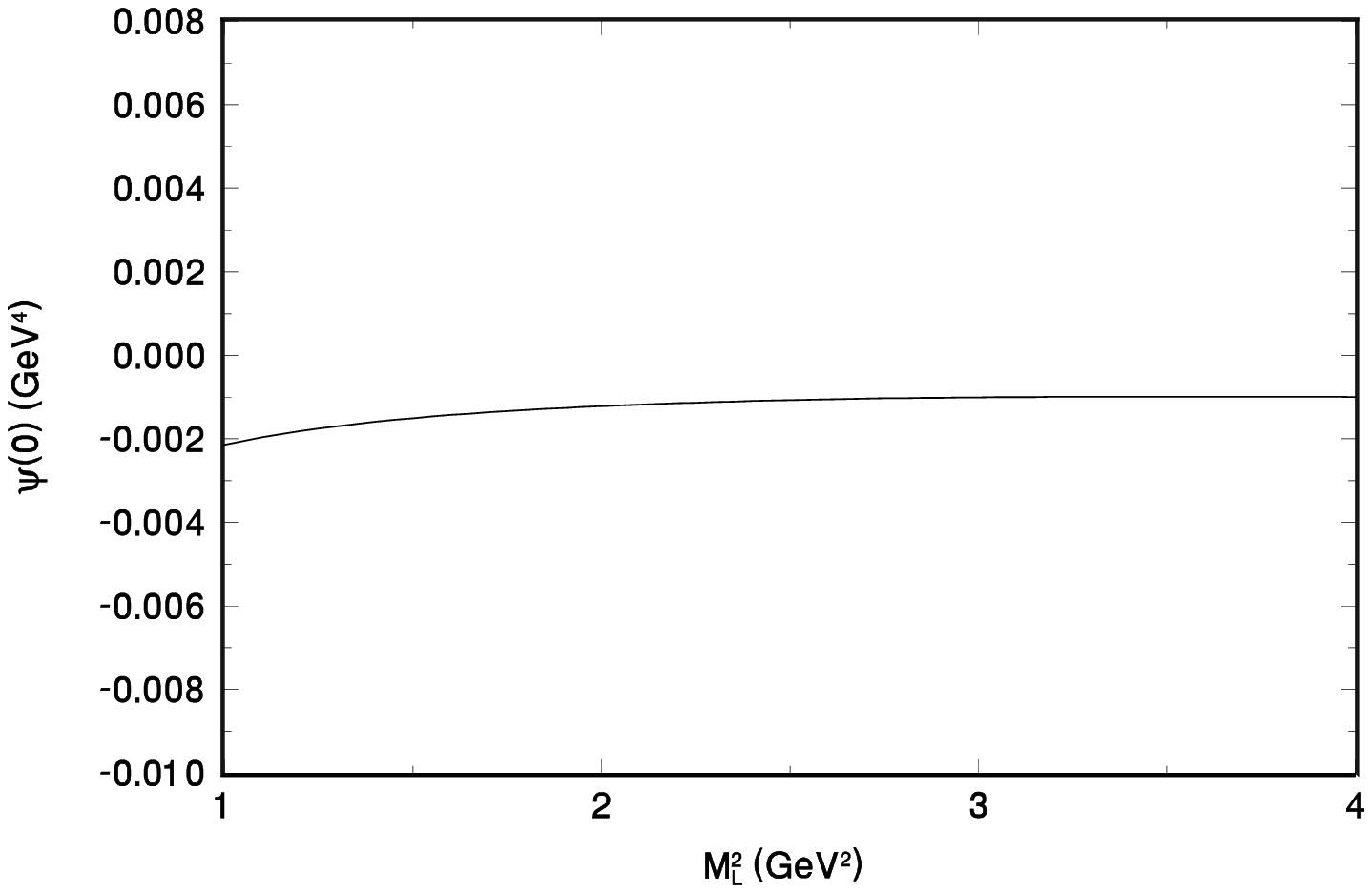}
\caption{}
\end{figure}
\begin{figure}[tp]
\epsffile{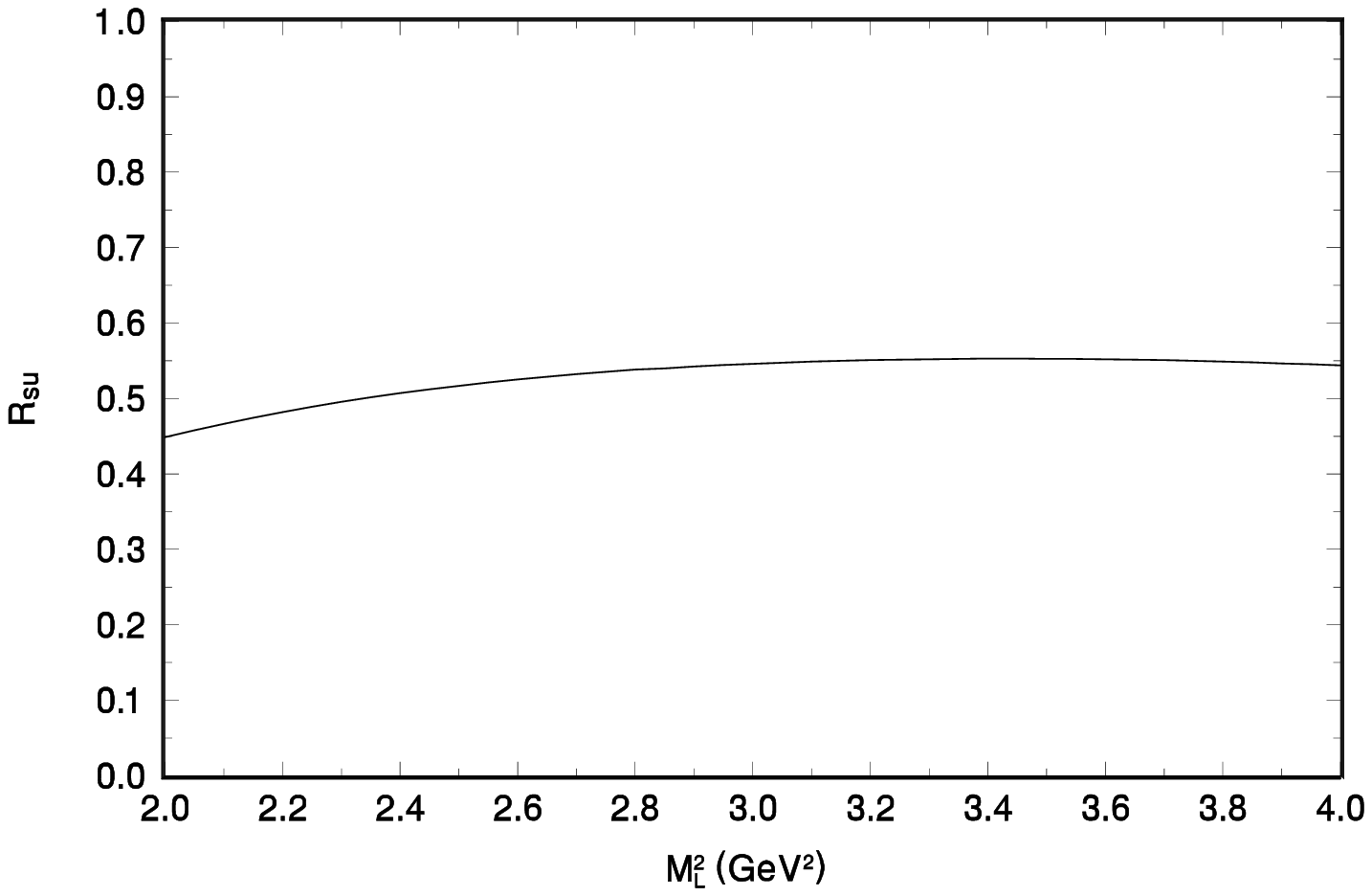}
\caption{}
\end{figure}

\end{document}